\documentclass[sprocl,12pt]{article}

\makeatletter
\def\bea{\begin{eqnarray}}
\def\eea{\end{eqnarray}}

\begin{document}

\title{ A generalized linear Hubble law for an
inhomogeneous barotropic Universe}
\author{J.-F. Pascual-S\'anchez\\
Dept. Matem\'atica Aplicada Fundamental, Secci\'on Facultad de
Ciencias,\\ Universidad de Valladolid, 47005, Valladolid,\\
Spain\\ E-mail: jfpascua@maf.uva.es}
\date{}
\maketitle

\footnotetext{Classical and Quantum Gravity in press}

\abstract{In this work, I present a generalized linear Hubble law
for a barotropic spherically symmetric
 inhomogeneous spacetime, which is in principle compatible with
 the acceleration of the cosmic expansion obtained as
  a result of high redshift Supernovae data.
The new Hubble function, defined by this law, has two additional
terms besides
 an expansion one, similar to the
 usual volume expansion one of the FLRW models, but now due to an
angular expansion.
 The first additional term is dipolar and is a consequence of the existence
 of a kinematic
  acceleration of the observer, generated by a negative
 gradient of pressure or of mass-energy density. The second one
 is quadrupolar and due to the shear. Both additional terms are
 anisotropic for off-centre observers,
 because of to their dependence on a telescopic angle of observation.
 This generalized linear
 Hubble law could explain, in a cosmological setting, the observed
large scale flow of matter, without to have recourse to peculiar
velocity-type newtonian models. It is pointed out also, that the
matter dipole direction should coincide with the CBR dipole one.}

\setlength{\parskip}{3cm}


\newpage

\setlength{\parskip}{4mm}

\section{Introduction}
 Last year [1], I have considered a specific inhomogeneous
 cosmological model, which could explain, in an alternative way,
 by the presence of
 a  kinematic acceleration generated by a
 negative gradient of pressure (or mass-energy), the present negative
  deceleration parameter, which appears to be a result of
 high redshift Supernovae data. These data were obtained by two different
 groups [2,3] and, if the SNe Ia are standard candles without
 evolution effects (although a recent work [4] suggests indicate
 that the SNe Ia are not standard candles), they show the existence
 of a cosmic acceleration, via the K-corrected redshift-luminosity
  distance relation. In the usual way, these data are explained, in the
   background of standard Friedmann
  models,
 by
  the presence of a positive cosmological constant or vacuum
  energy or quintessence.

This specific inhomogeneous spacetime, which I will also consider
 in this work, is a  barotropic (B), spherically (S)
  symmetric (hereafter BS). The comoving
matter congruence, in the vorticity-free BS inhomogeneous model,
has expansion, kinematic acceleration and shear, at difference
with the standard FLRW models, where it has expansion only.

At first sight, it seems that spherically symmetric inhomogeneous
models are non generic among the totality of inhomogeneous ones,
but note however that they can appear as a natural consequence of
inflation (see [5]). Note also that the measured almost isotropy
of the cosmic background radiation (CBR) temperature is in
principle compatible with large shear [6] or nonzero kinematic
acceleration [7], although these possibilities have yet to be
tested against the full range of cosmological observations,
including CBR polarization.

As the Cosmological Principle (CP) is not verified in the BS model
and, as usually the expansion linear Hubble's law is derived from
the CP, one can ask how this law, which characterizes the matter
flow, is generalized by the additional presence of acceleration
and shear in the BS model. Fortunately, the general linear Hubble
law, valid for any inhomogeneous Universe, is at our disposal. For
instance, it was presented in [8,9] and its expression, to linear
order in the angular diameter distance, is:
\begin{equation}\label{1}
z =\left(\frac{\theta}{3}-
\dot{u}_a\,o^a+\sigma_{ab}\,o^ao^b\right)\,D
\end{equation}
where $z$ is the measured redshift, $D$ is the angular diameter
distance to the emitter (however, note that, at linear order, the
Hubble law can be expressed in terms of any cosmological distance,
see [16]), $\theta$ the expansion, $\dot u_a$ is the 4-kinematic
acceleration of a comoving observer, $\sigma_{ab}$ is the shear
tensor and $o^a$, is an unit vector defined in observer's rest
space, which points in the observed direction of a light ray and,
hence, is opposite to the motion sense of the ray's photons.

This formula directly shows the isotropic, direccional and fully
anisotropic contributions to the redshift from the expansion,
acceleration and shear, respectively. All the terms between
parenthesis can be interpreted as a generalized Hubble function.
The first term, due to the cosmological expansion, gives the usual
expansion linear Hubble law of the FLRW models, which is sometimes
reinterpreted at small distances as a radial Doppler effect.
Although, we are  opposed to a newtonian kinematic
re-interpretation  at large distances of the cosmological
expansion redshift, however, it is clear that the expansion
redshift can be modified by nearly constant Doppler and
gravitational contributions, at both ends (and also in the light
travel) of the emission-observation process.

\section{The BS model}
 In order to specialize the general
formula (\ref{1}) to the BS model, we need to summarize its main
characteristics.
 We assume from the beginning that the cosmological
matter fluid is perfect and a mixture (radiation, baryonic matter,
neutrinos, etc.) and hence is characterized by a non-dust perfect
fluid congruence of unit 4-velocity $u^a$.
 In the BS model, the isotropy group
is 1-dim, whereas the isometry group is a 3-dim $G_3$, which is
acting multiply transitively on spacelike 2-dim surfaces
orthogonal to a preferred direction (hereafter $e_1$). These 2-dim
surfaces orbits, orthogonal both to the 4-velocity of the
congruence and to $e_1$, have constant Gaussian curvature but, of
course, they can be spheres, hyperboloids or planes. We consider,
in the BS model, the case of spherical symmetry.

 Also, in the BS model
 the vorticity is zero. This implies both the
existence of a global cosmic time and of 3-dim global spacelike
hypersurfaces orthogonal to the fluid congruence. In the BS model,
 the preferred spacelike direction $e_1$ is not
only geodesic and shear free in the local rest spaces orthogonal
to the 4-velocity $u^a$, moreover $e_1$ is geodesic along the
matter flow lines.

 Thus, the metric in comoving coordinates has the
expression
\begin{equation}
ds^2=-N^2(r,t)\, dt^2+B^2(r,t)\, dr^2+R^2(r,t)\, d\Omega^2 .
\end{equation}
where $d\Omega^2$ is the spherical line element and the
coefficient $N(r,t)$ is a lapse function which relates global
cosmic and local proper times. So, we have adopted in the 1+3
threading approach the Lagrangian formalism, with the comoving
identification, but the matter observers are Eulerian (due to the
null vorticity) in the ADM  3+1 slicing formalism.

 In the BS spacetime exists a preferred central worldline,
 i.e.
  the spacelike hypersurfaces have a
 centre at $r = 0$, where the isotropy group is 3-dim,
 and a preferred radial direction $e_1$ at each point,
associated with the direction of the only non-null component, $A$,
of the kinematic acceleration of the matter fluid elements. This
kinematic acceleration satisfies the spatially contracted
equations of motion
\begin{equation}\label{4}
 p^{\prime}+ (\mu +p)~A=0,
\end{equation}
where a prime denotes the derivative along the preferred radial
direction with unit vector field $e_1$. As we assume a barotropic
equation of state, $p(r,t) = p (\mu (r,t))$, with $\mu(r,t)$ the
mass-energy density, the last equation may also be written as:
\begin{equation}\label{mu}
c_s^2\,\mu ^{\prime}+ (\mu +p)\,A=0,
\end{equation}
where $c_s$ is the sound velocity. We see from the last formulas
that the kinematic acceleration, which opposes to the
gravitational attraction towards the centre, is generated by a
negative gradient of pressure or, equivalently, by a negative
gradient of mass-energy density. Note (see [1]), that the presence
in the metric of the coefficient $N(r,t)$, gives rise to a non
null kinematic acceleration, defined in the next section, and also
to a new term in the expression of the deceleration parameter $q$,
which was called in [1], inhomogeneity parameter, and was defined
there as
\begin{equation}\label{11}
I\!\!I= \frac{\alpha(t)}{(S\,H)^2},
\end{equation}
where $\alpha(t)$ is the coefficient of the second order term in
the radial expansion of the metric coefficient $N(r,t)$, and where
$S$ and $H$, are the scale expansion factor and  the usual
expansion Hubble function of FLRW models, respectively.

Depending on the value of this additional inhomogeneity term, the
deceleration  parameter $q$ can be positive or negative at present
cosmic time. In the last case, one has an alternative explanation
for the Supernovae data, in the realm of the BS model, as it has
been commented before.

\section{Generalized Hubble law for the BS model}
Now, we specialize the general Hubble law (\ref{1}), for the BS
model considered. Besides the null vorticity, the remaining
kinematical quantities have the following characteristics: the
scalar expansion $\theta(r,t)$ is a function of a radial comoving
coordinate and cosmic time, the 4-acceleration of the non-geodesic
comoving observers is $\dot u_a = (0, A(r,t), 0, 0)$, and the
shear mixed tensor has the diagonal coordinate expression
$\sigma_a^b = {\rm diag} (0,
\frac{2}{\sqrt{3}}\sigma(r,t),-\frac{1}{\sqrt{3}}
\sigma(r,t),-\frac{1}{\sqrt{3}} \sigma(r,t))$, where $\sigma^2
=\frac{1}{2}\sigma _a^b\sigma_b^a$.

The temporal component of both acceleration and shear is zero by
the choice of the comoving frame. On the other hand, the
4-acceleration has only one non-zero component in the sense of the
preferred spatial direction $e_1$ away from the centre, and the
shear mixed tensor the above diagonal form. Finally, due to
spherical symmetry, their components are only functions of $r$ and
cosmic time $t$. The expressions of the scalars $\theta, A$ and
$\sigma$, are
\begin{equation}
\theta(r,t) = \frac{\dot{B}}{B}+2\frac{\dot{R}}{R},
\end{equation}
\begin{equation}
A(r,t) =\frac{N^{\prime}}{N}
\end{equation}
and
\begin{equation}
\sigma(r,t) = \frac{1}{\sqrt{3}} \left (\frac{\dot{B}}{B}-
\frac{\dot{R}}{R}\right)       ,
\end{equation}
where the point means the proper time derivative, $\displaystyle
\frac{1}{N} \,\frac{\partial}{\partial t}$. So, the scalar shear
$\sigma$ is, except by a constant factor, the difference between
the radial and azimuthal expansion rates.

Now, performing a similar computation to that realized in [10] for
a dust subcase (Tolman-Bondi) of the BS model, the final
expression that adopts the linear Hubble's law in the BS model,
for off-centre observers $P_0$, is:
\begin{equation}\label{8}
    z = \left( \frac{\dot{R}}{R}-A\cos \Psi+
\sqrt{3}\,\sigma\cos ^2 \Psi\right)_0\, D
\end{equation}
By spherical symmetry, just the telescopic angle $\Psi$ is needed
to describe off-centre observations in the rest space of the
observer. $\Psi$ is the angle between the direction of observation
of a light ray, $o^a$, and the vector $e_1$, pointing radially
outward from the centre in the spacelike hypersurfaces of the BS
model.

 The generalized
Hubble function has three terms which contribute to the
cosmological redshift. The first is similar to the usual one of
FLRW models due to the volume expansion, but in this model is due
to the azimuthal expansion and depends not only on cosmic time but
also on a radial comoving distance or position of the observer
with respect to the centre. The azimuthal expansion comes from the
expansion of the geometrical $S^2$ sphere centered at the observer
($r = r_0$) in the spacelike hipersurface ($t = t_0$). Thus, the
azimuthal expansion implies a variation in cosmic time of the
angle $\Psi$, under which a specific emitter is observed in the
observational celestial sphere of the rest space of the observer.
The second term in the generalized Hubble function is a dipolar
one, due to the acceleration (which does not appear in the
Tolman-Bondi subcase), and a third quadrupolar one, due to the
shear.

 The inhomogeneous BS model has a preferred central worldline
 (a spatial centre in the
 spacelike hypersurfaces)
  in which the mass-energy density and the
pressure are global maxima at any cosmic time, and where
 the acceleration and the shear are zero. Hence, at the
centre, the matter congruence and CBR are exactly isotropic, but
for off-centre observers as us, additional terms appear due to the
radial inhomogeneity of this spacetime.

The consequences of the generalized linear Hubble law are
striking. For instance, when we are observing in the same sense
that the acceleration away from the centre, that is $\Psi=0$, then
the acceleration term gives a maximum violetshift contribution,
$-AD$, due to the fact that we are expelled out from the centre by
the kinematic acceleration. Of course, observing in the opposite
sense to the acceleration, i.e., towards the centre, it gives an
maximum dipole redshift, $+AD$. In both cases, if the emitters are
at the same distance, the additional expansion and shear
quadrupole terms have the same positive value. Hence, using the
new Hubble law (\ref{8}), the difference of redshifts of these
kind of observations, is a pure dipole violetshift, which is only
due to the kinematic acceleration $A$. Therefore, we consider this
specific direction away from the centre, as the global direction
of the matter dipole.

Moreover, as it have been stated before, the dipole term can also
be modified by nearly constant dipole redshifts originated by
Doppler and gravitational contributions in the process of
emission-observation of light. By using the BS model, these
additional redshifts must be calculated as deviations from the
generalized Hubble law (\ref{8}) valid for comoving non-geodesic
observers. However, as it has been stated above, for isolating the
pure dipole part, it is necessary to make the difference between
redshift observations of emitters situated at the same distance,
but in opposite senses from the observer.

Additional dipole voiletshifts are normally interpreted, by using
Newton's theory, as deviations from an isotropic linear Hubble
expansion law due to peculiar velocity fields or a bulk flow
towards a local "Great Attractor", (see [11]).

However, in the model considered, the centre of the spacelike
3-dim hypersurfaces is the "global repulsor" and is the centre of
the rest-frame of matter and also of the rest-frame of the CBR.
Hence the matter dipole direction given by the generalized Hubble
law must coincide with the CBR dipole direction. With respect to
the CBR dipole, we refer to a recent work [12], where all the
anisotropies including non-linear effects have been calculated,
and where it was estimated, how the kinematic acceleration
contributes to the dipole temperature anisotropy and the shear to
the quadrupole one. I leave for future work the obtention of the
CBR dipole and quadrupole temperatures in the BS model and the
influence of the new form of the Hubble function on its
deceleration parameter.

\section{Final comments and conclusions}
 Other works which have
tried to explain the matter or the CBR dipole as a cosmological
effect (see for instance [10,13,14]), have considered a matter
congruence of geodesic observers, because all use the dust
Tolman-Bondi spacetime (except the recent work [15] on a special
Stephani spacetime). Instead, in the BS model, the gradient of
pressure is not zero, and hence, the congruence of matter
observers is non-geodesic, except at the centre, in which it is
supposed to be the origin of the matter flow. Note that, even in
the late Universe, when the pressure is small (if the mean
velocity of Galaxies is around $300$ Km/sec or $10^{-3}\, c$, then
the pressure is around $10^{-6´}$ of the mass-energy density),
however, the gradient of pressure could not be necessarily small.

The main difference between the cosmological matter dipole of the
new law (\ref{8}) and the usual dipole, manifested as a deviation
of the isotropic Hubble law of Friedmann models, is that the
latter is interpreted as a constant Doppler effect, i.e.
independent of the emitter's distance. This Doppler dipole arises
in the standard model from peculiar velocities, and it can be
eliminated going to the "correct" zero peculiar velocity frame.
Whereas, the cosmological acceleration dipole, measured off the
centre in the fundamental comoving frame of non-geodesic matter
observers, is emitter's distance dependent and cannot be
eliminated for off-centre observers. The same happens for the
quadrupole shear term.

However, we do not claim that all the observed matter dipole is
cosmological. In our model, a peculiar velocity dipole induced by
a local inhomogeneity, must be calculated as a local perturbation
of the inhomogeneous background spacetime, i.e. as a local
deviation of the new Hubble law (\ref{8}).

Finally note, that the new Hubble function of this model is not
only cosmic time dependent, as in FLRW models, but observer's
position and angular dependent too. Therefore,
 this may account
for the difference between its inferred values from observations
performed with different telescopic angles.

\begin{center}
{\large Acknowledgements}
\end{center}
\vspace*{-.3cm} I am grateful to A. San Miguel and F. Vicente for
many discussions on this and (un)related subjects. Special thanks
are due to a referee, for helping me to improve the presentation
of the  manuscript. This work is partially supported by the
spanish research projects VA34/99 of Junta de Castilla y Le\'on
and C.I.C.Y.T. PB97-0487.

\begin{center}
{\large REFERENCES}
\end{center}
\vspace*{-.3cm} [1] Pascual-S\'anchez, J.-F., {\it Mod. Phys.
Lett. A} {\bf 14}, 1539 (1999).\\ \noindent [2] Perlmutter, S., et
al., {\it Astrophys. J.} {\bf 517},
 565 (1999).\\
\noindent [3] Riess, A.G., et al., {\it Astronomical J.} {\bf
116}, 1009 (1999).\\
 \noindent
 [4] Riess, A. G., preprint astro-ph/9907038, (1999).\\
 \noindent
   [5] Linde, A., Linde, D., and Mezhlumian, A., {\it Phys. Lett. B} {\bf
345}, 203 (1995).\\
 \noindent
 [6] Nilsson, U.S., Uggla, C., Wainwright, J., Lim, W.C., {\it Astrophys. J. Lett.}\\ \hspace*{.4cm}
  {\bf 521}, L1 (1999).\\
 \noindent
[7] Ferrando, J.J., Morales, J.A., Portilla, M., {\it Phys. Rev.
D} {\bf 46}, 578 (1992).\\
 \noindent
 [8] Ehlers, J., {\it Gen. Rel. Grav.}
{\bf 25}, 1225 (1993), english translation of\\ \hspace*{.4cm}
{\it Akad. Wiss. Lit. Mainz, Abhandl Math.-Nat. Kl.} {\bf 11}, 793
(1961).\\
 \noindent
 [9] Ellis, G.F.R., in {\sl 'General
Relativity and Cosmology'}, ed. R.K. Sachs (N.Y.:\\
 \hspace*{.5cm} Academic Press) (1971).\\
\hspace*{-.1in}
  [10] Humphreys, N. P., Maartens, R., and Matravers, R., {\it Astrophys.
J.} {\bf 477}, 47 (1997).\\
 \hspace*{-.1in}
 [11] Hudson, M.J., preprint astro-ph/9908036, (1999).\\
 \hspace*{-.1in}
  [12] Maartens, R., Gebbie, T., Ellis, G.F.R., preprint
astro-ph/9808163v2, (1999).\\
 \hspace*{-.1in}
 [13] Paczynski, B., Piran, T., {\it Astrophys. J.} {\bf 364}, 341 (1990).\\
\hspace*{-.1in}
 [14] Langlois, D., Piran, T., {\it Phys. Rev. D}
{\bf 53}, 2908 (1996).\\
 \hspace*{-.1in}
  [15] Clarkson, C.A., Barrett, R.K., {\it Class. Quantum Grav.} {\bf 16}, 3781
(1999).\\
 \hspace*{-.1in}
  [16] Hasse, W., Perlick, V., {\it Class. Quantum Grav.} {\bf 16}, 2559 (1999).

\end{document}